\documentclass{article}
\usepackage{emulateapj}
\usepackage{apjfonts}

\newcommand{\cm}{\rm cm} 
\newcommand{\yr}{{\rm yr}}
\newcommand{\kms}{{\rm km}\,{\rm s}^{-1}}
\newcommand{\K}{{\rm K}}
\newcommand{\kpc}{{\rm kpc}}
\newcommand{\Mpc}{{\rm Mpc}}
\newcommand{\Msun}{{{\rm M}_\odot}}

\newcommand{\HI}{\ion{H}{1}} 
\newcommand{\MgI}{\ion{Mg}{1}} 
\newcommand{\MgII}{\ion{Mg}{2}} 
\newcommand{\NaI}{\ion{Na}{1}} 
\newcommand{\FeII}{\ion{Fe}{2}} 

\newcommand{\lya}{Ly$\alpha$} 

\lefthead{Schaye}
\righthead{High-redshift galaxies and DLA systems}

\begin{document}

\submitted{2001, ApJ, 559, L1}

\title{\small On the relation between high-redshift starburst galaxies and
damped Ly$\alpha$ systems} 
\author{Joop~Schaye}
\affil{School of Natural Sciences, Institute for Advanced
Study, Einstein Drive, Princeton NJ 08540, schaye@ias.edu}

\begin{abstract}
Essentially all high-redshift galaxies show evidence for strong
large-scale outflows that should have a profound effect on the
structure and kinematics of both the galaxies themselves and their
environment. The interstellar absorption spectra of both Lyman break
galaxies (LBGs) and local starburst galaxies are remarkably similar to
those of damped \lya\ (DLA) systems in QSO spectra. Their rest-frame
UV spectra typically show broad, blueshifted \lya\ absorption
accompanied by low-ionization metallic absorption lines with complex
profiles. Hydrodynamical simulations of galactic winds suggest that
the absorption arises in a collection of dense clouds entrained in the
hot wind and/or in shells swept up by the outflow. It seems likely
that outflows would also give rise to DLA absorption when seen in
absorption against a background QSO.  It is emphasized that some
differences are expected between the properties of DLAs in the
emission-weighted galaxy spectra, which always probe the centers of
star formation, and those in random sight lines through the
outflow. The observed LBGs alone can account for all DLA absorption at
$z\sim 3$ if the cross section for DLA absorption is $\pi r^2$ with $r
= 19~h^{-1}~\kpc$. If the cross section is smaller than this, then the
fraction of DLAs arising in outflows could still be significant if
there are many wind-driving galaxies below the current detection
limits. This is certainly possible since the $z=3$ luminosity function
is still rising down to the detection limit of $0.1~L_\ast$ and
observations and simulations of local dwarf galaxies indicate that
many of these fainter galaxies drive winds as well.
\end{abstract}


\section{Introduction}
Damped Ly$\alpha$ (DLA) systems give rise to the highest column
density absorption lines found in the spectra of distant quasars. By
definition, DLA systems have neutral hydrogen column densities greater than
$2\times 10^{20} ~\cm^{-2}$, large enough for the line profiles to
show prominent damping wings due to the natural width of the \lya\
transition. Despite years of intense study, the nature of high-redshift
DLA systems remains uncertain, primarily because DLA systems are generally too faint
to be detected in emission with current instrumentation. 

In recent years, inferences from the internal kinematics of DLA systems,
traced by the associated heavy-element absorption lines from
low-ionization species, have been the most controversial. The velocity 
profiles are generally asymmetric and show multiple narrow components
distributed over 10s to 100s km/s (e.g., Prochaska \& Wolfe 1997,
1998). There are two types of models that have
been claimed to reproduce the observed kinematics. Prochaska \&
Wolfe (1997, 1998) tested a variety of idealized models and concluded
that, of these, only rapidly rotating, thick disks can reproduce the
observations. This would be very problematic for the cold dark matter
(CDM) model which generically predicts that the cross section for DLA
absorption is dominated by systems with low rotation speeds (e.g.,
Kauffmann 1996). In contrast, Haehnelt, Steinmetz, \& Rauch (1998) analyzed a
hydrodynamical simulation of a CDM model and concluded that merging
protogalactic subclumps can match the observed kinematics (but see
McDonald \& Miralda-Escud\'e 1999).

Although some low-$z$ DLA systems are known to be massive spiral galaxies
(e.g., le Brun et al.\ 1997), the hypothesis that a significant
fraction of high-$z$ DLA systems 
arise in rapidly rotating disks conflicts with several observations.
If high-redshift DLA systems were massive spirals, then it would be difficult
to explain why they are not detected in deep imaging studies, why
their metallicities are so low [$-2.0 \la \log(Z/Z_\odot) \la -0.5$;
e.g., Pettini et al.\ 1994] and why they have such
low molecular fractions (typically $< 10^{-6}$, Petitjean, Srianand,
\& Ledoux 2000). Furthermore, all high-$z$ DLA systems detected in 21cm 
absorption have much higher spin temperatures than local spiral
galaxies (e.g., Carrili et al.\ 1996; Chengalur \& Kanekar 2000).

Current state-of-the-art cosmological simulations of DLA systems do not
include a number of important physical processes, such as cooling by
molecules and metals, thermal and dynamical effects of
shelf-shielding, and feedback from star formation. Furthermore, the
simulation output is corrected for self-shielding using highly
simplified prescriptions, which can lead to large errors in the
neutral hydrogen column densities, and the numerical resolution is
still too low to resolve the relevant spatial and mass scales. Given
these caveats, it would perhaps be surprising if cosmological
simulations were able to reproduce both the kinematics and the column
density distribution of DLA systems. Haehnelt et al.\ (1998) find that they
can reproduce the kinematics, but did not investigate the column
density distribution. Others (e.g., Katz et al.\ 1996) have only
studied the latter and found that their simulations 
produced too few self-shielded absorbers. In short, although
cosmological simulations are invaluable tools, current results
regarding the nature of DLA systems may not be robust.

Here it is argued that recent observations of high-redshift
galaxies and local starbursts, as well as hydrodynamical simulations,
suggest that galactic winds may give rise to the DLA systems seen in the
spectra of QSOs. The idea that the complex kinematic structure of DLA systems
is due to feedback from star formation was first discussed by York et
al.\ (1986), who suggested that the internal motions due to ongoing
star formation in Magellanic-type ``dwarf'' galaxies are
responsible. More recently, Nulsen, Barcons, \& Fabian (1998) argued
that massive outflows from dwarf galaxies give rise to DLA absorption
and showed that, in their semi-analytic model for galaxy formation,
outflows from dwarf galaxies are common enough to account for the
majority of DLA systems. This letter is complementary to the paper of
Nulsen et 
al.\ insofar as neither observations of high-redshift galaxies and
local starbursts nor hydrodynamical simulations were discussed in the
latter.

\section{DLA systems and outflows}
\subsection{Observational evidence}

Observations show that at high redshift ($z \sim 2-4$) there is a
large population of 
vigorously star forming galaxies which differ from present-day spiral
and elliptical galaxies. These so-called Lyman break galaxies (LBGs)
are compact, with half-light radii of a few
kiloparsecs\footnote{Unless stated 
otherwise, all distances are physical.} (e.g., Giavalisco, Steidel, \&
Macchetto 1996; Lowenthal et al.\ 1997) and have high star formation rates
(SFR $\sim 1$ -- $100~\Msun\,\yr^{-1}$; Steidel et al.\ 1996; Lowenthal
et al.\ 1997). Both their rest-frame UV spectra (e.g., Steidel et
al.\ 1996; Ebbels et al.\ 1996; Lowenthal et al.\ 1997; Franx et al.\
1997; Pettini et al.\ 2000, 2001) and ultraviolet-to-optical spectral
energy distributions (Papovich, Dickinson, \& Ferguson 2001) are
remarkably similar to those of local starburst galaxies (e.g., Conti,
Leitherer, \& Vacca 1996; Gonz\'ales Delgado et al.\ 1998; Kunth et
al.\ 1998). Essentially all $z\sim 3$ LBGs show evidence for
large-scale outflows, with velocities of $100 - 1000~\kms$ (e.g.,
Pettini et al.\ 1998, 2001). Observations of local dwarf galaxies
indicate that outflows are also common in low-mass galaxies with star
formation rates $10^{-3}- 10^{-1}~\Msun\,\yr^{-1}$ (e.g., Heckman et
al.\ 1995, 2001b), i.e., orders of magnitude lower than in the
observable LBGs. 

Although the statistics of \lya\ absorption in LBGs has not yet been
investigated, it is clear that a large fraction of LBGs show
blueshifted DLA absorption. In particular, the average LBG
spectra of Lowenthal et al.\ (1997) and Steidel, Pettini, \&
Adelberger (2001) show 
DLA absorption. Ly$\alpha$ emission is generally weak or
absent. The \lya\ absorption is accompanied by strong, blueshifted
low-ionization interstellar metal lines with complex profiles and a
large velocity dispersion (e.g., Pettini et al.\ 2000). The
interstellar absorption spectra of LBGs (and local starbursts) are
thus similar to those of DLA systems in QSO spectra. 

However, as emphasized by Pettini et al.\ (2000), there may also be
some differences. For example, the velocity dispersion of the
low-ionization metal lines in the spectrum of the LBG MS 1512-cB58 is
somewhat higher ($\sim 500~\kms$) than is typical of
DLA systems. Furthermore, the presence of fine-structure lines, which are
rarely detected in DLA systems, suggests that the gas density may be
higher. Pettini et al.\ (2001) used nebular emission lines to measure
the oxygen abundance for a sample of bright LBGs and estimate $\left
[{\rm O}/{\rm H}\right ] \ga -1.0$ with an uncertainty of about 1
dex. Although these results are uncertain, they do suggest that bright
LBGs may have metallicities somewhat greater than those typical of
DLA systems.

It is important to note that systematic differences between DLA systems in
random sight lines to background sources and DLA systems in the spectra of
the starbursting galaxies are unavoidable, even if all DLA systems were to
arise in outflows. One reason is that emission-weighted spectra probe
the inner regions of the starburst, which are denser, are more enriched,
and have a larger velocity dispersion than the outer part of the
outflow, which dominates the cross section for interception by a
random sight line. In fact, depending on the orientation of the
galaxy, a sight line through the outflow may not intersect the galaxy
at all. The line of sight velocity dispersion will also vary with the
orientation of the sight line relative to the outflow.  With regard to
the metallicities it is also important to note that the abundances
derived from nebular emission lines cannot be directly compared to
abundances derived from interstellar absorption lines, as the latter
could include a large contribution from metal-poor swept-up
interstellar and intergalactic gas. Finally, as will be discussed
below, it is likely that the cross section for DLA absorption is
dominated by galaxies of lower luminosity, and therefore probably also
metallicity, than the few LBGs that are bright enough to do
spectroscopy. 

In an important study, Heckman et al.\ (2000) used \NaI\ absorption
lines to investigate outflows from local, far-IR bright starburst
galaxies. They found that for the majority of galaxies the
interstellar Na~D line was blueshifted relative to the systemic
velocity by over 100~$\kms$. From the measured equivalent widths of the
\NaI\ lines, Heckman et al.\ estimate that the total gas column
density $N_H$ in the flow is approximately a few times $10^{21}~\cm^{-2}$,
i.e., column densities typical of DLA systems.  Hence, the observations of
Heckman et al.\ suggest that, if seen in absorption against a
background QSO, the outflows would give rise to a DLA system accompanied by a
complex system of low-ionization metal lines spanning a velocity range
of up to hundreds of kilometres per second.

A direct way of testing the hypothesis that outflows from starbursting
dwarf galaxies give rise to DLA absorption is to analyze spectra of
QSOs at various impact parameters from known starburst
galaxies. Unfortunately, starbursts and quasars that are bright enough
to do spectroscopy are rare, and consequently there are few known pairs
with small separations.  Norman et al.\ (1996) searched for absorption
by low-ionization gas in sight lines to QSOs located close to the two
local starburst galaxies NGC~520 and NGC~253. Even though the impact
parameters of all sight lines were large ($\ge 24\,h^{-1}~\kpc$), \MgI,
\FeII\ and strong \MgII\ were detected. From the complex structure of
the \MgII\ absorption and the fact that it arises beyond the last
21~cm emission contours (corresponding to $N_{HI} = 2\times 10^{19}
~\cm^{-2}$), the authors concluded that the absorption arises in dense
cloudlets that are too small to be resolved by 21~cm maps even though
they probably have large \HI\ column densities.

\subsection{Theoretical considerations}
\label{sec:theory}

Some idealized calculations of the absorption lines arising in
galactic outflows have been published (e.g., Wang 1995; Theuns, Mo, \&
Schaye 2001), but the results are uncertain due to the unknown 
structure of the winds and because thermal and hydrodynamical
instabilities are not taken into account. I will therefore focus on
results from hydrodynamical simulations. However, it has to
be noted that the results from these simulations are also
uncertain, given that they are generally two-dimensional, include only
some of the relevant physics, consider only isolated galaxies embedded
in an intergalactic medium (IGM) without substructure, and follow only
the initial stages of the starburst.

Analytic and numerical modeling (e.g., Suchkov et al.\ 1994; Silich \&
Tenorio-Tagle 1998; Strickland \&  
Stevens 1999; Mac Low \& Ferrara 1999; Murakami \& Babul 1999;
D'Ercole \& Brighenti 1999) has lead to the following picture of the
starburst phenomenon. Correlated supernova explosions and winds from
massive stars inject a large amount of energy into the insterstellar
medium (ISM), thereby
generating a hot, expanding bubble which sweeps up a cold ($T \la
10^4~\K$) shell of interstellar material. As the bubble blows out of
the disk, the expanding bubble accelerates down the density gradient
and the shell fragments via Rayleigh-Taylor (RT) instabilities. The
hot gas flows out between the fragments sweeping up new shells. The
velocity field is complex and generally consists 
of a central (bi)conical outflow surrounded by large-scale
vortices. Behind the shock, a two-phase medium arises. Cold cloudlets
are formed via RT and thermal instabilities and destroyed by turbulent
mixing, thermal conduction and evaporation. Dense fragments are
accelerated by the ram pressure of the wind and are thus susceptible
to the Kelvin-Helmholtz instability, unless they are self-gravitating.
At large distances from the galaxy, swept-up
shells are decelerating and therefore RT-stable, but eventually some
may fall back onto the galaxy and again become RT-unstable.

Although current simulations cannot capture all the complex physics,
it is clear that a large amount of cold, dense gas spanning a wide
range of velocities is present in the outflow. Heckman et al.\ (2000)
analyzed a hydrodynamical simulation of M82's galactic wind and find
that within $\left | z \right | \le 1.5~\kpc$ of the disk the cold
material covers a projected area of $\sim 2 ~\kpc^2$ at $N_H \sim
9\times 10^{21}~\cm^{-2}$; for $\left | z \right | > 1.5~\kpc$ the
average column density is $N_H\sim 10^{20}~\cm^{-2}$ for a projected
area of $8.5~\kpc^2$. Hence, it is likely that a sight line to a
background QSO that passes through the outflowing region would contain
a DLA. The cloudlets and fragments would then give rise to
discrete low-ionization metal lines which could have a large 
velocity dispersion.

Efstathiou (2000) showed that efficient feedback is also possible in a
quiescent mode of star formation. A dwarf galaxy can expel a large
fraction of its gas over a time-scale $\sim 1~$Gyr with a star
formation rate that never exceeds $0.1~\Msun \,\yr^{-1}$. Mac Low \&
Ferrara (1999) showed that winds from dwarf galaxies (with ${\rm SFR}
\sim 10^{-4} - 10^{-3}~\Msun\,\yr^{-1}$) can eject a large fraction of
the metals without strongly disturbing the ISM of the galaxy. These
findings suggest that although most dwarf galaxies should be driving
winds, these winds need not destroy the disk. Thus, both the disk and
the outflow could contribute to the DLA absorption, and the latter
could explain the large velocity dispersion of the metal lines.
However, it is unclear whether the high rates of star formation and
the high merger rates of high-redshift galaxies allow the presence of
a stable, thin disk.  Furthermore, even if winds from dwarf galaxies
do not disturb a large fraction of their ISM, winds from neighboring
galaxies could.

\section{The rate of incidence}

In the previous sections, it was argued that galactic winds can give
rise to DLA absorption similar to that seen in the spectra of
QSOs. The key question is then: Are galactic winds common enough to
account for a large fraction of high-redshift DLA systems? To answer this
question, one needs to know the number density of wind-driving galaxies
and the cross section for DLA absorption per galaxy, both of which are,
unfortunately, poorly constrained.

Given the luminosity function of high-redshift galaxies, it is
straightforward to estimate the typical cross section for DLA
absorption per galaxy required to account for all DLA systems. Let $n$ be the
comoving number density of galaxies and let each galaxy provide a
physical cross section $\pi r^2$ for DLA absorption. Neglecting
overlap between galaxies, the total rate of incidence of DLA
absorption is,
\begin{eqnarray}
{dN \over dz} &=& n\pi r^2 {c \over H(z)} (1+z)^2 \nonumber \\
&=&  0.20 \left ({ r \over 19 ~h^{-1}~\kpc}\right )^2
\left ( n \over 0.016~h^3\,\Mpc^{-3}\right ),
\label{eq:dn/dz}
\end{eqnarray}
for $(\Omega_m,\Omega_\Lambda) = (0.3,0.7)$ and $z=3$.
Storrie-Lombardi \& Wolfe (2000) measure $dN/dz = 0.20$ at $z\sim
3$. By combining their own survey with data from the Hubble Deep
Field, Steidel et al.\ (1999) find that the comoving number density of
observed LBGs at $z\sim 3$ is $n=0.016 ~h^3\,\Mpc^{-3}$ down to an
$\mathcal{R}$ magnitude of 27.0, which corresponds to $0.1
L_\ast$. Hence, equation (\ref{eq:dn/dz}) implies that if every observed
galaxy provides a physical cross section for DLA absorption of $\pi
r^2$ with $r = 19 ~h^{-1}\,\kpc$ (3.5 arcsec), then they could account
for all DLA systems observed in the spectra of QSOs and a typical DLA
counterpart would be significantly fainter than $L_\ast$.

The fact that spectra of LBGs show strong \lya\ absorption implies
that typically a region at least as large as the light emitting region
must be covered\footnote{It is unclear whether the absorption lines in
the spectra of LBGs are black in the centers. This implies that,
although the covering factor of the high column density gas must
clearly be large, it may be less than unity (see Heckman et al.\ 2001a
for some discussion).} by a large neutral hydrogen column. Bright LBGs
are observed to have compact cores with half-light radii of 1 -- 7 kpc
(Giavalisco et al.\ 1996; Lowenthal et al.\ 1997), which are often
surrounded by lower surface brightness nebulosities. Assuming that
every LBG spectrum contains a DLA, this yields a lower limit to the
fraction of DLA systems associated with observed LBGs, which are known to
drive winds, of 0.3 -- 14 $h^2$ percent.

The fraction of DLA systems associated with outflows could, of course, be much
greater if there are wind-driving galaxies below the Hubble Deep Field
detection limit. For example, if the luminosity function measured by
Steidel et al.\ (1999) continues down to $0.01 L_\ast$, then $n=0.095
~h^3\,\Mpc^{-3}$, which would imply $r = 7.9 ~h^{-1}\,\kpc$ (1.5
arcsec). Hydrodynamical simulations suggest that the number of
galaxies is greater than this. For example, Theuns et al.\ (2001) find
$n \approx 0.5 ~h^3\,\Mpc^{-3}$ for stellar masses greater than $1.3
\times 10^7~\Msun$ at $z=3$ which would imply $r = 3.4 ~h^{-1}\,\kpc$
(0.63 arcsec).

Hence, regardless of whether DLA systems arise in galactic winds, their
counterparts are probably predominantly faint ($L \ll L_\ast$) and
have small impact parameters ($\la 1$ arcsec). More sophisticated
models (e.g., Fynbo, M{\o}ller, \& Warren 1999) yield similar
results. This picture 
is consistent with current observations. Attempts to detect high-$z$
DLA counterparts in emission have established that associated galaxies
must typically be significantly fainter than $L_\ast$ and/or close to
the line of sight ($\la 1$ arcsec) (e.g., Steidel, Pettini, \&
Hamilton 1995; Warren et al.\ 2001). 

In short, a large fraction of high-$z$ DLA systems could arise in galactic
winds if a substantial fraction of $L \ll L_\ast$ galaxies drive
large-scale outflows.

\section{Discussion and conclusions}

Observations of LBGs show that at high redshift there is a large
population of compact galaxies which have high star formation rates
and drive large-scale outflows. The spectra of LBGs and local
starburst galaxies typically show blueshifted DLA absorption and
associated low-ionization metallic absorption lines with complex
profiles. Thus, the interstellar absorption spectra of LBGs and local
starburst galaxies resemble those of DLA systems in QSO spectra, implying
that the outflows themselves are responsible for at least a subset of
DLA systems. Some systematic differences are, however, expected. Sight
lines to 
background QSOs are more likely to intersect the outer parts of the
galaxy/outflow, whereas emission-weighted galaxy spectra probe the
inner parts. Hence, DLA systems in the spectra of galaxies may arise in gas
that is denser, is more enriched, and has a higher velocity than is
typical of DLA systems in QSO spectra. Furthermore, spectroscopy is currently
only possible for the brightest LBGs, which may have higher
metallicities than the fainter galaxies which dominate the
cross section for DLA absorption.

The observed LBGs alone can account for all $z\sim 3$ DLA absorption
detected in the spectra of QSOs if each galaxy provides a physical
cross section for DLA absorption of $\pi r^2$ with $r =
19~h^{-1}~\kpc$, considerably larger than the typical half-light radii
of LBGs. If the cross section for DLA absorption is smaller, DLA systems
arising in outflows could still dominate if there are many
wind-driving galaxies that are too faint to detect with current deep
imaging studies. This is certainly possible since the $z=3$ luminosity
function is still rising down to the detection limit of $0.1~L_\ast$,
and observations of local dwarf galaxies suggest that even galaxies
with star formation rates many orders of magnitude lower than is
typical of observed LBGs ($10^{-3} - 10^{-1} ~\Msun\,\yr^{-1}$ vs.\
$10^0 - 10^2 ~\Msun\,\yr^{-1}$) drive winds.

Hydrodynamical simulations suggest that DLA absorption in the spectra
of starburst galaxies arises in dense cloudlets entrained in the hot
wind and/or in (fragments of) shells swept up by the outflow. Although
I have focused on DLA systems, multiphase outflows will, of course, also
give rise to absorption lines of lower column densities, including
Lyman limit and \MgII\ systems.

Aguirre et al.\ (2001) predict that winds from observed LBGs escape
the galaxies' potentials and propagate to large distances ($\sim 10^2
~\kpc$ in about 1 Gyr). Clearly, the covering factor for DLA
absorption cannot be close to unity at such large radii or winds would
greatly overproduce the observed number of DLA systems. Observations indicate
that at least the inner few kpc are covered by neutral hydrogen
columns typical of DLA systems. As the radius increases, the wind's ram and
thermal pressure drops and consequently the column densities of
entrained pressure-confined clouds decreases. Self-gravitating clouds
may survive, but their covering factor will decrease as one over the
radius squared. Hence, the expectation is that the sphere of influence
of the winds that escape the galaxy extends far beyond the maximum
impact parameter for DLA absorption.  At large radii the \lya\
absorption will in fact be greatly reduced, leading to gaps in the
\lya\ forest, if the winds sweep up a significant fraction of the IGM
and/or shock-heat the IGM (Theuns et al.\ 2001).

Feedback from star formation is generally thought to play a major role
in the formation and evolution of galaxies. If galactic winds are
important, then one would expect to see their signature in DLA systems
since these are the highest column density absorption lines
detected. Recent observations and simulations suggest that galactic
winds give rise to the DLA phenomenon if seen in absorption against a
background quasar. Whether outflows provide a large fraction of the
total cross section for high-redshift DLA absorption is still
unclear. If winds are important, then this would complicate the
interpretation of observations of DLA systems, but it would also mean
that the large amount of high-quality observations of DLA systems can
be used to study one of the most important unsolved problems in galaxy
formation.

\acknowledgments
I am grateful for useful discussions with Anthony Aguirre, Max
Pettini, Chuck Steidel, and Tom Theuns. 
This work was supported by a grant from the W. M. Keck foundation.

\end{document}